\documentclass[10pt,prl,twocolumn,showpacs,superscriptaddress,floatfix]{revtex4-2}
\usepackage{graphicx}
\usepackage{amssymb}
\usepackage{fontspec}
\usepackage{amsmath}
\usepackage{amsthm}
\usepackage{bm}
\usepackage{physics}
\usepackage{color,xcolor}
\usepackage{subfigure}
\usepackage{algpseudocode}
\usepackage{soul}
\usepackage{algorithm}
\usepackage{algorithmicx}
\usepackage{lineno}
\usepackage{soul,xcolor}
\usepackage{color}
\usepackage{mathtools} 
\usepackage{mathrsfs}
\usepackage{graphicx}
\usepackage{dcolumn}
\usepackage{bm}
\usepackage{graphicx}
\usepackage{textcomp,mathcomp}
\usepackage{enumerate}
\newtheorem{theorem}{Theorem}

\newtheorem{corollary}{Corollary}
\usepackage{lipsum} 

\floatname{algorithm}{Algorithm} 

\usepackage[colorlinks,
linkcolor=blue,      
anchorcolor=blue, 
citecolor=purple]{hyperref}

\begin{document}

\preprint{APS/123-QED}

\title{Noise-Resilient Heisenberg-limited Quantum Sensing via Indefinite-Causal-Order Error Correction}

\author{Hang Xu}
\affiliation{State Key Laboratory of Photonics and Communications, Institute for Quantum Sensing and Information Processing, Shanghai Jiao Tong University, Shanghai 200240, People’s Republic of China}%

\author{Xiaoyang Deng}
\affiliation{State Key Laboratory of Photonics and Communications, Institute for Quantum Sensing and Information Processing, Shanghai Jiao Tong University, Shanghai 200240, People’s Republic of China}%
\author{Ze Zheng}
\affiliation{State Key Laboratory of Photonics and Communications, Institute for Quantum Sensing and Information Processing, Shanghai Jiao Tong University, Shanghai 200240, People’s Republic of China}%


 \author{Tailong Xiao}%
\email{tailong\_shaw@sjtu.edu.cn}
\affiliation{State Key Laboratory of Photonics and Communications, Institute for Quantum Sensing and Information Processing, Shanghai Jiao Tong University, Shanghai 200240, People’s Republic of China}%
 \affiliation{Hefei National Laboratory, Hefei, 230088, People’s Republic of China}
\affiliation{Shanghai Research Center for Quantum Sciences, Shanghai, 201315, People’s Republic of China}

 \author{Guihua Zeng}%
 \email{ghzeng@sjtu.edu.cn}
\affiliation{State Key Laboratory of Photonics and Communications, Institute for Quantum Sensing and Information Processing, Shanghai Jiao Tong University, Shanghai 200240, People’s Republic of China}%
 \affiliation{Hefei National Laboratory, Hefei, 230088, People’s Republic of China}
\affiliation{Shanghai Research Center for Quantum Sciences, Shanghai, 201315, People’s Republic of China}

\date{\today}

\begin{abstract}
Quantum resources can, in principle, enable Heisenberg-limited (HL) sensing, yet no-go theorems imply that HL scaling is generically unattainable in realistic noisy devices. While quantum error correction (QEC) can suppress noise, its use in quantum sensing is constrained by stringent requirements, including prior noise characterization, restrictive signal--noise compatibility conditions, and measurement-based syndrome extraction with global control. Here we introduce an ICO-based QEC protocol, providing the first application of indefinite causal order (ICO) to QEC. By coherently placing auxiliary controls and noisy evolution in an indefinite causal order, the resulting noncommutative interference enables an auxiliary system to herald and correct errors in real time, thereby circumventing the limitations of conventional QEC and restoring HL scaling. We rigorously establish the protocol for single- and multi-noise scenarios and demonstrate its performance in single-qubit, many-body, and continuous-variable platforms. We further identify regimes in which error correction can be implemented entirely by unitary control, without measurements. Our results reveal ICO as a powerful resource for metrological QEC and provide a broadly applicable framework for noise-resilient quantum information processing.
\end{abstract}

\maketitle


\textit{Introduction.}---Quantum resources, such as coherence and entanglement, allow quantum sensors to reach the Heisenberg limit (HL) \cite{HL1,HL2,HL3,HL4,HL5,HL6}. However, environmental noise is inevitable in practical settings \cite{noisysensing0,noisysensing1,noisysensing2,noisysensing3,noisysensing4}. A rigorous no-go theorem dictates that, even with maximally entangled states, noisy sensors cannot sustain HL scaling \cite{nogo, zss1}. While quantum error correction (QEC) offers a pathway to circumvent this theorem \cite{qec1,qec2,qec3,qec4,qec5,qec6,qec7,qec8,qec9}, current protocols face significant implementation hurdles. Repetition codes \cite{qec4}, for instance, require spatially parallel sensing channels—a resource that is often unavailable. Alternatively, stabilizer codes utilizing noise-free ancillas avoid direct signal encoding on auxiliary qubits, yet they strictly require prior knowledge of the noise type and satisfaction of the Hamiltonian-not-in-Lindblad-span (HNLS) condition \cite{qec5}. For a single-qubit sensor, this implies the severe constraint that only a single noise channel orthogonal to the signal Hamiltonian can exist. Furthermore, if experimental capabilities are limited to local Pauli measurements, the syndrome extraction in these protocols would destructively interfere with the signal encoding.


Theories of quantum gravity \cite{gravity1,gravity2,gravity3,gravity4} suggest that spacetime geometries may exist in superposition, implying that the causal order of events can be indefinite. This concept has been formalized as indefinite causal order (ICO) \cite{ICO1,ICO2,ICO3}, with the quantum SWITCH serving as a paradigmatic realization \cite{SWITCH1,SWITCH2}. While ICO offers established advantages in quantum computation \cite{ico_comput1,ico_comput2,ico_comput3} and communication \cite{ico_commun1,ico_commun2}, its application to metrology reveals a complex landscape. In continuous-variable systems, ICO enables the estimation of time-dependent geometric phases with super-Heisenberg precision \cite{ico_cvsensing1,ico_cvsensing2}; conversely, for time-independent unitary channels, the precision remains strictly bounded by the Heisenberg limit (HL) \cite{ico_dvsensing1,ico_dvsensing2,ico_dvsensing3}. This dichotomy motivates a pivotal question: can ICO be harnessed to mitigate noise—effectively functioning as a QEC mechanism—to enhance estimation precision in open quantum systems?

In this Letter, we introduce an indefinite-causality quantum error–correction (IQEC) protocol that circumvents the no-go theorems for noisy quantum sensors, representing the first use of ICO for QEC design. Our protocol employs a set of auxiliary gates whose causal order with the noisy sensing channel is coherently controlled by a quantum SWITCH. The resulting non-commutative interference drives the auxiliary system into orthogonal states, enabling error syndromes to be extracted solely through measurements on the auxiliary space. We describe how to construct the indefinite-causality channels, characterize the admissible set of auxiliary gates, and show that suitable choices exist for all Pauli noise models. The protocol’s performance is further validated in representative physical systems. Remarkably, for phase-covariant noise \cite{phase-covariant}, syndrome measurements become unnecessary: purely unitary control suffices. Together, these results highlight four advantages of our approach: all Pauli noise can be corrected without prior information, error diagnosis requires only auxiliary-space measurements, the signal Hamiltonian faces no structural constraints from the noise, and in certain scenarios the protocol operates without any measurement and real-time feedback at all.

\textit{Noisy quantum sensing.}---
In quantum sensing, the estimation precision of a parameter $\omega$ is given by the variance ${\mathop{\rm Var}}(\hat \omega )$ of its estimator $\hat \omega$, and the lower bound of the variance is defined by the quantum Cramér-Rao bound \cite{qfi1,qfi2}
\begin{equation}
{\mathop{\rm Var}}(\hat \omega )  \ge {M^{ - 1}}{\cal F}_Q^{ - 1},\label{inequality}
\end{equation}
where $M$ is the number of shots and ${\cal F}_Q$ is quantum Fisher information (QFI).
The QFI provides the ultimate bound for the variance of the estimation via symmetric logarithmic derivatives \cite{qfi1}.

For a closed system, the sensing precision can achieve the HL (${\cal F}_Q \propto {t^2}$).
However, practical sensors are inevitably plagued by environmental noise, and the quantum advantage is strictly confined to the coherence time. 
Beyond this period, the sensitivity degrades to the standard quantum limit (${\cal F}_Q \propto {t}$).
While QEC is pivotal for quantum computing, its application in sensing is complicated by the inherent tension between error suppression and signal accumulation.
Consider a probe qubit with Hamiltonian $\omega \sigma^P_z$ coupled to an ancilla. A standard stabilizer code may define the code space as $\mathscr{C} = \text{span}\{\ket{0}_A\ket{0}_P, \ket{1}_A\ket{1}_P\}$ and the error space as $\mathscr{E} = \text{span}\{\ket{0}_A\ket{1}_P, \ket{1}_A\ket{0}_P\}$. 
In this case, the stabilizer measurement can detect transverse noise $\sigma^P_x$ without disturbing the signal accumulation. 
However, parallel noise $\sigma^P_z$ remains undetectable.

\begin{figure}[htbp]
  \centering
\includegraphics[width=0.9\linewidth]{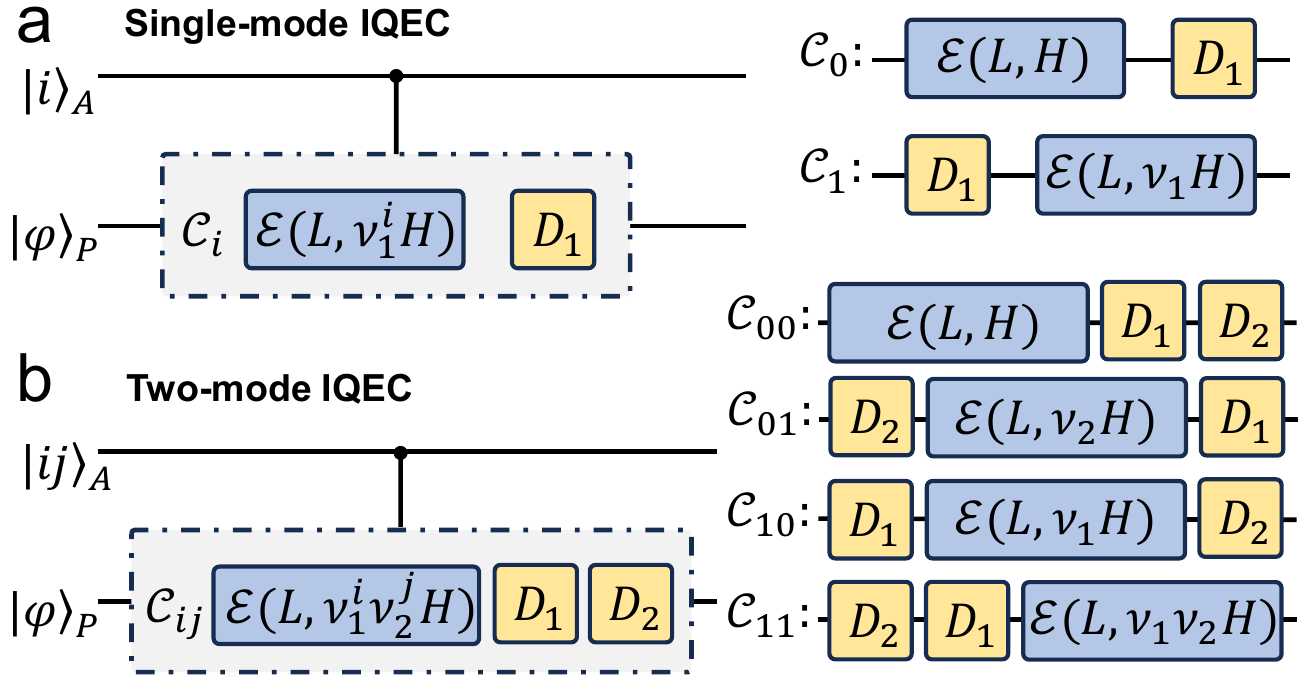}
  \caption{Schematic of the IQEC protocol. 
  The evolution order ${\cal{C}}_{ij}$ of the probe is controlled by the computational basis of the auxiliary qubits.}
  \label{frame}
\end{figure}

\textit{The IQEC protocol.}---We work in the short-time regime and describe the probe evolution by an effective quantum channel accurate to first order in time.
Specifically, for a small interrogation time $t$, we approximate the reduced dynamics by a CPTP map
\begin{equation}
\mathcal{E}_{L,H} \circ \rho_P
=K_0(t)\rho_P K_0^\dagger(t)+\sum_i K_i(t)\rho_P K_i^\dagger(t)+O(t^2),
\label{eq:short_channel}
\end{equation}
where $U(t)=e^{-iHt}$ and
\begin{equation}
K_i(t)=\sqrt{q_i(t)}\,L_iU(t),\;
K_0(t)=\Big(\mathbb I-\frac12\sum_i q_i(t)\,L_i^\dagger L_i\Big)U(t).
\label{eq:kraus_short}
\end{equation}
Here $q_i(t)$ are $O(t)$ coefficients determined by the underlying reduced dynamics (e.g., $q_i(t)=\int_0^t\kappa_i(s)\,ds$ when a time-local generator exists).
We denote $L=\{L_1, L_2,\cdots\}$ as the noise set.
Eq.~\eqref{eq:short_channel} neglects $O(t^2)$ corrections, including higher-order terms mixing the unitary and dissipative contributions.

Firstly, we apply ICO-based QEC to the single-noise setting with jump operator $L=L_1$ \cite{L}. We introduce an auxiliary qubit $A$ and initialize the composite system $S=A\otimes P$ in $\ket{+}_A\ket{\varphi}_P$. We further apply a noiseless auxiliary gate $D_1$. In the ICO protocol, the auxiliary state coherently controls the causal order: conditioned on $\ket{0}_A$ the probe undergoes the sensing evolution followed by $D_1$, whereas conditioned on $\ket{1}_A$ the order is reversed. We first assume $[D_1,H]=0$. In the no-jump trajectory, the final state reads
\begin{equation}
\begin{split}
\ket{f}_S &= \frac{1}{\sqrt{2}} \left[ \ket{0}_A D_1 e^{-iHt}\ket{\varphi}_P + \ket{1}_A e^{-iHt} D_1 \ket{\varphi}_P \right] \\
&= \ket{+}_A \otimes D_1 e^{-iHt} \ket{\varphi}_P.
\end{split}
\end{equation}
Conversely, if a quantum jump occurs (generated by $L_1$), the state becomes
\begin{equation}
\ket{f}_S = \frac{1}{\sqrt{2}} \left[ \ket{0}_A D_1 L_1 e^{-iHt}\ket{\varphi}_P + \ket{1}_A L_1 e^{-iHt} D_1 \ket{\varphi}_P \right].
\end{equation}
Choosing $D_1$ such that it anticommutes with the noise, $\{D_1,L_1\}=0$, yields the error-tagged state
\begin{equation}
\ket{f}_S = \ket{-}_A \otimes D_1 L_1 e^{-iHt} \ket{\varphi}_P.
\end{equation}
A measurement of the auxiliary qubit in the $\sigma_x$ basis therefore heralds the jump without erasing the accumulated signal, enabling perfect correction.

The constraint on $D_1$ can be further relaxed by incorporating an indefinite time direction (ITD)~\cite{ITD1,ITD2,ITD3,ITD4,ITD5,ITD6}, a concept motivated by quantum-gravity frameworks. Here we consider the case $\{D_1,H\}=0$. The auxiliary qubit then controls not only the causal order but also the time direction of the Hamiltonian evolution: $\ket{0}_A$ selects forward dynamics $\mathcal{E}_{L,H}$, while $\ket{1}_A$ selects backward dynamics $\mathcal{E}_{L,-H}$. In the absence of a jump, the final state is
\begin{equation}
\begin{split}
\ket{f}_S &= \frac{1}{\sqrt{2}} \left[ \ket{0}_A D_1 e^{-iHt}\ket{\varphi}_P + \ket{1}_A e^{iHt} D_1 \ket{\varphi}_P \right] \\
&= \ket{+}_A \otimes D_1 e^{-iHt} \ket{\varphi}_P,
\end{split}
\end{equation}
where we used $\{D_1,H\}=0$ to rewrite $e^{iHt}D_1=D_1e^{-iHt}$. If a jump occurs, one obtains
\begin{equation}
\ket{f}_S = \ket{-}_A \otimes D_1 L_1 e^{-iHt} \ket{\varphi}_P,
\end{equation}
which is identical to the commutation case above. Hence, ICO provides the essential operational framework for the IQEC protocol, while invoking ITD is optional and depends on the algebraic compatibility between $D_1$ and $H$.

We formalize these processes via the superchannel $\mathcal{S}(D,\mathcal{E}_{L,H})$. Fig.~\ref{frame} provides a schematic of the IQEC protocol, where $\nu_i = H D_i / D_i H$ characterizes the time-reversal action induced by $D_i$. The operating mode of IQEC is set by the number of auxiliary qubits. Explicit superchannel constructions for arbitrary-mode IQEC are given in the Supplemental Material (SM). The above analysis extends directly to multiple-noise scenarios, yielding the following theorem (proofs in SM):
\begin{theorem}\label{the1}
Consider a set of noises $L \coloneqq \{L_i\}_{i=1}^m$ of the probe. For a minimal group-generating set $G \coloneqq \{ L'_i \}_{i=1}^{m_a}$ of $L$, if there exists a set of gates $D \coloneqq \{ D_i\}_{i=1}^{m_a}$ satisfying
\begin{equation}
\label{c2}
\begin{aligned}
    &\{ D_i, D_j \} \quad\mathrm{or}\quad [D_i, D_j] = 0, \quad i \ne j; \\
    &\{ D_i, H \} \quad\mathrm{or}\quad [D_i, H] = 0; \\
    &\{ D_i, L'_i \} = 0; \quad [D_i, L'_j] = 0, \quad i \ne j,
\end{aligned}
\end{equation}
then the IQEC protocol only requires $m_a$ auxiliary qubits to correct the noise set.
\end{theorem}
Here any element in $L$ can be represented as a product of elements from $G$:
\begin{equation}
\begin{array}{l}
\forall {L_i} \in L, \;\; {L'_k} \in G,\\
{L_i} = c{L'_j} \quad \text{or} \quad c{L'_j}{L'_k} \quad \text{or} \quad c{L'_j}{L'_k}{L'_m} \cdots,
\end{array}
\end{equation}
where $c$ is an arbitrary constant. 
Note that this minimal generating set is not unique \cite{G_eg} and may sometimes coincide with the noise set itself. 

\begin{figure}[htbp]
  \centering
  \includegraphics[width=1\linewidth]{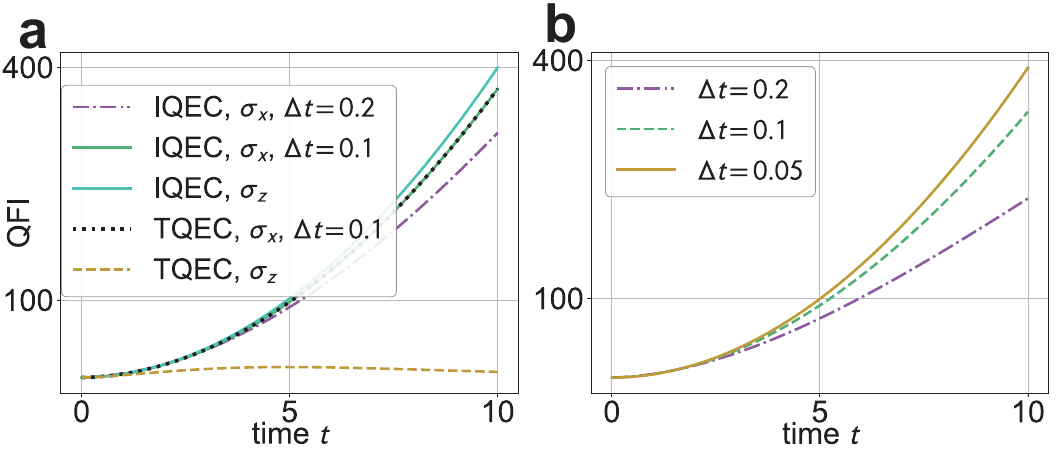}
  \caption{
  (a) Results of the single qubit for different protocols under single noise. The $\sigma_x$ or $\sigma_z$ in the legend denotes the noise operator. 
  (b) Results of the single qubit for IQEC protocol under unknown noises. }
  \label{1bitqfi}
\end{figure}

\textit{Single qubit.}---Consider a single-qubit sensor governed by the Hamiltonian $H=\omega\sigma_z$ and subject to a single noise source. When the noise operator is orthogonal to the Hamiltonian, traditional QEC (TQEC) can correct the resulting errors. IQEC can also correct such orthogonal noise by choosing $D=\sigma_z$; however, it still requires an ultrashort correction interval $\Delta t$ to mitigate the noncommutativity, and therefore offers no practical advantage over TQEC. By contrast, for noise parallel to the Hamiltonian, $L=\sigma_z$, TQEC fails altogether. In this case, IQEC can satisfy Theorem~\ref{the1} by choosing $D=\sigma_x$. Moreover, since the noise commutes with the Hamiltonian, error correction is only needed at the final time. Fig.~\ref{1bitqfi}(a) shows the IQEC performance for the single-noise case.

We next consider the multi-noise setting $\{\sigma_x,\sigma_y,\sigma_z\}$. Since the auxiliary-gate generator $G$ can be chosen from $\{\sigma_x,\sigma_y\}$, we introduce two auxiliary qubits and two auxiliary gates $\{\sigma_y,\sigma_x\}$. The joint system is initialized in $\ket{++}_A\ket{\varphi}_P$. After the evolution through the superchannel ${\cal S}(D,{\cal E}_{L,H})$, each occurrence of a jump operator $L_i$ flips a designated auxiliary qubit, thereby encoding the corresponding error event into an orthogonal auxiliary state.

We further analyze IQEC in the presence of unknown noise. A conservative strategy is to assume that all relevant error mechanisms may occur, including Pauli errors and amplitude damping. Since any single-qubit noise can be expanded in the operator basis $\{I,\sigma_x,\sigma_y,\sigma_z\}$, the sensing channel can be written in the generalized Kraus form
\begin{equation}
{\cal E}_{L,H}\circ \ket{\varphi}_P\!\bra{\varphi}
= \sum_{i,j\in\{I,x,y,z\}}
c_{ij}\,
\sigma_i e^{-iHt} \ket{\varphi}_P\!\bra{\varphi}\, e^{iHt} \sigma_j^\dagger ,
\label{gKc}
\end{equation}
where $\sigma_I=\mathbb{I}$ and $c_{ij}$ are expansion coefficients.

Passing through the superchannel ${\cal S}$, each Kraus operator $\sigma_i$ maps the auxiliary system to a unique orthogonal state. Consequently, measuring the auxiliary system removes the off-diagonal contributions ($i\neq j$ in Eq.~(\ref{gKc})) and yields an effective mixture of standard Kraus channels for the probe:
\begin{equation}
\begin{split}
& \mathrm{Tr}_A\!\left[
{\cal S}(D,{\cal E}_{L,H})\circ
\ket{\varphi}_P\!\bra{\varphi}
\right] \\
& = \sum_{i\in\{I,x,y,z\}}
c_i\, D_i\, \sigma_i e^{-iHt}
\ket{\varphi}_P\!\bra{\varphi}\,
e^{iHt} \sigma_i^\dagger D_i^\dagger ,
\end{split}
\label{sKc}
\end{equation}
where each branch is calibrated according to the auxiliary measurement outcome (see SM). Fig.~\ref{1bitqfi}(b) demonstrates that, within the short-time approximation regime, IQEC restores HL scaling even when Pauli noise and amplitude damping coexist, confirming its applicability to unknown-noise environments.

\begin{figure}[htbp]
  \centering
  \includegraphics[width=1\linewidth]{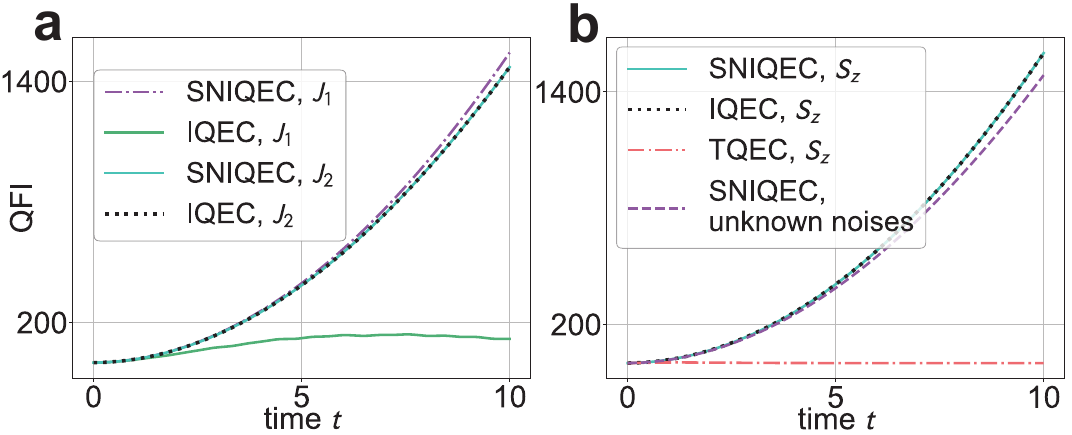}
  \caption{Results of the two-qubit for different protocols with different noises.}
  \label{2bitqfi}
\end{figure}

\textit{Many-body system.}---For many-body systems, we assume a locally generated Hamiltonian,
$H=\sum_{k=1}^N H^{(k)}$.
Because spatially separated sites can be controlled independently, the time-evolution directions associated with $H^{(k)}$ need not be uniform across the lattice (see SM).
We refer to this architecture as the \emph{spatially non-uniform} SNIQEC protocol.
We begin by restricting attention to local and correlated noise processes in the set $\mathcal{W}$.

\begin{theorem}\label{the2}
For an $N$-site system with Hamiltonian $H=\sum_{k=1}^N H^{(k)}$, consider a noise set $L \coloneqq \{L_i\}_{i=1}^m$ acting on the probe, where $L_i\in\mathcal{W}$.
Let $G_L \coloneqq \{ {L'}_i^{(k)}\}_{i=1}^{m_b^{(k)}}$ be a minimal \emph{local} group-generating set for $L$.
If there exists a collection of gates $D \coloneqq \{D_i^{(k)}\}_{i=1}^{m_b^{(k)}}$ such that
\begin{equation}
\label{c3}
\begin{aligned}
    &\{D_i^{(k)},D_j^{(k)}\}=0 \quad\mathrm{or}\quad [D_i^{(k)},D_j^{(k)}]=0, \quad i\neq j; \\
    &\{D_i^{(k)},H^{(k)}\}=0 \quad\mathrm{or}\quad [D_i^{(k)},H^{(k)}]=0; \\
    &\{D_i^{(k)},{L'}_i^{(k)}\}=0; \qquad [D_i^{(k)},{L'}_j^{(k)}]=0,\quad i\neq j,
\end{aligned}
\end{equation}
then the SNIQEC protocol requires only $\sum_{k=1}^N m_b^{(k)}$ auxiliary qubits to correct the noise set.
\end{theorem}

The minimal local group-generating set is defined analogously to $G$, with the additional constraint that each generator ${L'}_i^{(k)}$ acts nontrivially only on its corresponding site $k$.
Importantly, for multi-qubit probes and any noise channel $L_i\in\mathcal{W}$, one can always construct a set of auxiliary gates that enables SNIQEC to correct all local and correlated noise.
By contrast, within the \emph{standard} IQEC architecture there need not exist any auxiliary-gate set that both satisfies Theorem~\ref{the1} and corrects the same noises.
A concrete example is the Hamiltonian $H=\omega(\sigma_z^{(1)}+\sigma_z^{(2)})$ with the noise set
$J_1=\{\sigma_\theta^{(1)},\,\sigma_\theta^{(2)},\,\sigma_\theta^{(1)}\sigma_\theta^{(2)}\}$ \cite{J1}.

The auxiliary-qubit overhead of an error-correcting scheme is set by the size of the (local) group-generating set.
While the size $m_a$ of $G$ satisfies $m_a\le m$, the local size $\sum_{k=1}^N m_b^{(k)}$ is not bounded in the same manner.
For instance, consider the two-qubit noise set
$J_2=\{\sigma_x^{(1)}\sigma_x^{(2)},\,\sigma_y^{(1)}\}$ \cite{J2}.
By Theorems~\ref{the1} and \ref{the2}, both IQEC and SNIQEC apply, but IQEC requires fewer auxiliary qubits and does not rely on site-resolved control of the evolution direction.
Fig.~\ref{2bitqfi}(a) shows the corresponding two-qubit results, consistent with the two noise scenarios discussed above.

In many-body settings, another important class is \emph{collective} noise of the form $\sum_{k=1}^N L^{(k)}$.
For spin systems, a canonical example is $S_z=\sum_{k=1}^N \sigma_z^{(k)}$.
For such processes, even when a jump is detected, there is generally no unitary gate that inverts the jump, since the collective operator $L$ is intrinsically irreversible.
Nevertheless, the evolution under collective noise admits a generalized Kraus representation.
Within the superchannel ${\cal S}(D,{\cal E}_{L,H})$, each local contribution $L^{(k)}$ maps the auxiliary system to a distinct orthogonal subspace; hence, correcting the local contributions effectively corrects the full collective operator $L$.
This reasoning extends directly to broader classes of collective noise.

\begin{corollary}
For an $N$-site system with Hamiltonian $H = \sum_{k=1}^N H^{(k)}$, consider a set of noises $L \coloneqq \{L_i\}_{i=1}^m$ of the probe, where $L_i \in \mathcal{K}$ (local, correlated, and collective noises set). 
For a minimal algebra-generating set $G_E \coloneqq \{{L'}_i^{(k)}\}_{i=1}^{m^{(k)}_c}$ of $L$, if there exists a set of gates $D \coloneqq \{D^{(k)}_i\}_{i=1}^{m_c^{(k)}}$ satisfying Eq.~(\ref{c3}),
then $\sum_{k=1}^N m_c^{(k)}$ auxiliary qubits are required to correct these noises via SNIQEC.
\end{corollary}

Here, any element of the noise set $L$ can be generated from the algebra-generating set via repeated products and linear combinations:
\begin{equation}
\begin{array}{l}
\forall {L_i} \in L, \;\; {{L'}_i^{(k)}}\in G_E,\\ 
{L_i} = c{L'}_i^{(k)}\;\;{\rm or}\;\;c{L'}_i^{(k)}{L'}_j^{(l)}\;\;{\rm or}\;\;c{L'}_i^{(k)} + c'{L'}_j^{(l)} \cdots .
\end{array}
\label{c4}
\end{equation}
where $c,c'\in\mathbb{C}$.
For multi-qubit probes, the algebra-generating set associated with any trace-preserving noise process can always be reduced to local Pauli operators $\{\sigma_x^{(k)},\sigma_y^{(k)}\}$.
These generators satisfy Eq.~(\ref{c4}), implying that SNIQEC can correct arbitrary trace-preserving noise on a many-body system.
Consequently, for an $N$-qubit probe, all noise events can be heralded using $2N$ auxiliary qubits, independent of the microscopic noise model.

SNIQEC is, however, unnecessary for purely collective noise.
As an example, consider the Hamiltonian $H=\omega S_z$ with collective noise $L=S_z$.
This model violates the HNLS condition, rendering TQEC inapplicable, yet IQEC remains feasible.
To attain maximal sensitivity, we initialize the probe in a GHZ state.
If a collective jump occurs, it can be undone by applying an appropriate recovery operator~\cite{UL}.

Fig.~\ref{2bitqfi}(b) compares the sensing performance of different error-correction strategies in a two-qubit system under collective noise and under unknown trace-preserving noise, confirming the advantage of our protocol in the collective-noise regime and its applicability to unknown noise processes.

\begin{figure}[htbp]
  \centering
  \includegraphics[width=1\linewidth]{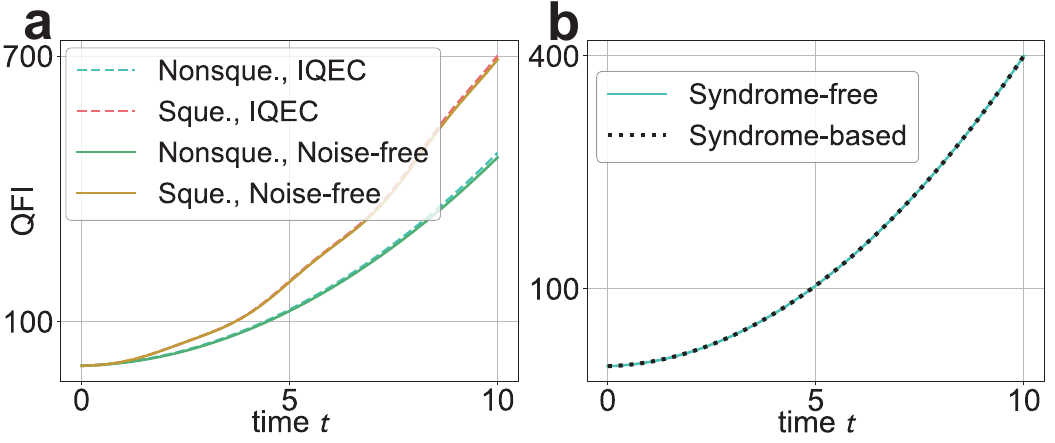}
  \caption{
  (a) Results of the continuous variable system, where $\Delta t=0.25$.
  (b) Results of the syndrome-based and
syndrome-free schemes for IQEC.
  }
  \label{cvqfi}
\end{figure}

\textit{Continuous variable.}---We consider a bosonic probe governed by
$H=\omega a^\dagger a+\chi\,(a^2+a^{\dagger 2})$
and subject to photon-loss noise $L=a$.
The probe is initialized in a cat state.
In the nonsqueezed regime~\cite{squeezed}, the evolution remains confined to the cat subspace, under which the action of $a$ is effectively reversible.
Invoking Theorem~\ref{the1} with the auxiliary parity gate $D=e^{i\pi\hat n}$, IQEC therefore achieves perfect error correction.
With squeezing present, the dynamics generally drive the state out of the cat manifold, rendering photon loss irreversible: IQEC can still herald loss events, but active recovery is no longer possible.
Nevertheless, in close analogy with weak-value protocols, the probe can be purified by post-selecting against the photon-loss trajectory.
Fig.~\ref{cvqfi}(a) shows that IQEC attains noise-free precision in the nonsqueezed case, while in the squeezed case the same precision is recovered via post-selection.


\textit{Unitary correction.}---When the noise operators commute with the Hamiltonian, error correction is required only at the end of the evolution, allowing a fully unitary implementation without intermediate measurements or syndromes.
\begin{theorem}\label{the3}
Consider a set of noises $L \coloneqq \{L_i\}_{i=1}^m$ of the probe. 
For a minimal algebra-generating set $G_F \coloneqq \{ L'_i \}_{i=1}^{m_d}$ of $L$, if there exists a set of gates $D \coloneqq \{ D_i\}_{i=1}^{m_d}$ satisfying satisfying Eq.~(\ref{c2}), and $[H,L]=0$, then with $m_d$ auxiliary qubits, IQEC can achieve error correction using only unitary control.
\end{theorem}
Here, $L'_i$ is unitary, ${L'_i}^2 =\pm \mathbb I$, and $L$ are phase-covariant noises \cite{PC}. 
For a single qubit, with the Hamiltonian $\omega \sigma_z$ and noise $\sigma_z$, Theorem~\ref{the3} holds. 
Fig.~\ref{cvqfi}(b) illustrates results for both syndrome-based and syndrome-free schemes, both of which restore the HL.

\textit{Conclusions.}---In summary, we have introduced an indefinite-causal-order–based QEC (IQEC) protocol that provides a general framework for restoring Heisenberg-limited (HL) sensitivity in noisy quantum sensors. By inserting a set of auxiliary gates whose action is placed in an uncertain order relative to the sensing dynamics, IQEC exploits the resulting non-commutativity to detect noise via measurements on the auxiliary system. We establish general constraints on the noise and auxiliary operations, and show that IQEC achieves correctability and recovers HL scaling for arbitrary Pauli noise—including irreversible channels—thereby ensuring applicability even when the noise is unknown. We further demonstrate the protocol’s efficacy across single-qubit, many-body, and continuous-variable platforms, where it reliably restores noise-free HL performance. Under specific conditions, IQEC simplifies substantially, requiring neither frequent corrections nor large ancillary systems, and in some cases eliminating the need for syndrome measurements altogether. 
These features relax the noise requirements for high-precision sensing and establish ICO as a valuable quantum resource for QEC. Beyond sensing, IQEC offers a versatile error-mitigation framework across NISQ-era systems \cite{nisq1}, with potential applications in QEC-enhanced computing \cite{correction1,correction2,computing1,computing2}, communication \cite{communication1,communication2}, and batteries \cite{battery1,battery2}.


\textit{Acknowledgments.}---The authors acknowledge support from the National Key R\&D Program of China (No.~2025YFF0515504), the National Natural Science Foundation of China (No.~62401359), the State Key Laboratory of Photonics and Communications, the Quantum Science and Technology—National Science and Technology Major Project (No.~2021ZD0300703), and the Shanghai Municipal Science and Technology Major Project (No.~2019SHZDZX01).

\nocite{*}

\bibliography{sample}



\end{document}